\documentclass[fleqn,10pt]{wlscirep}
\title{Spin polarized semimagnetic exciton-polariton condensate in magnetic field}

\author[1,*]{Mateusz~Kr\'ol}
\author[1]{Rafa\l~Mirek}
\author[1]{Katarzyna~Lekenta}
\author[1]{Jean-Guy~Rousset}
\author[1]{Daniel~Stephan}
\author[1]{Micha\l~Nawrocki}
\author[2]{Micha\l~Matuszewski}
\author[1]{Jacek~Szczytko}
\author[1]{Wojciech~Pacuski}
\author[1]{Barbara~Pi\k{e}tka}

\affil[1]{Institute of Experimental Physics, Faculty of Physics, University of Warsaw, ul.~Pasteura 5, PL-02-093 Warsaw, Poland}
\affil[2]{Institute of Physics, Polish Academy of Sciences, al.~Lotnik\'{o}w 32/46, PL-02-668 Warsaw, Poland}

\affil[*]{mateusz.krol@fuw.edu.pl}

\usepackage[T1]{fontenc}
\usepackage{graphicx}
\usepackage{dcolumn}
\usepackage{bm}
\usepackage{amsmath}
\usepackage{lmodern}
\usepackage{mathtools}

\begin{document}

\begin{abstract}

Owing to their integer spin, exciton-polaritons in microcavities can be used for observation of non-equilibrium Bose-Einstein condensation in solid state. However, spin-related phenomena of such condensates are difficult to explore due to the relatively small Zeeman effect of standard semiconductor microcavity systems and the strong tendency to sustain an equal population of two spin components, which precludes the observation of condensates with a well defined spin projection along the axis of the system. The enhancement of the Zeeman splitting can be achieved by introducing magnetic ions to the quantum wells, and consequently forming semimagnetic polaritons. In this system, increasing magnetic field can induce polariton condensation at constant excitation power. Here we evidence the spin polarization of a semimagnetic polaritons condensate exhibiting a circularly polarized emission over 95\% even in a moderate magnetic field of about 3~T. Furthermore, we show that unlike nonmagnetic polaritons, an increase on excitation power results in an increase of the semimagnetic polaritons condensate spin polarization. These properties open new possibilities for testing theoretically predicted phenomena of spin polarized condensate. 
\end{abstract}

\thispagestyle{empty}
	\flushbottom
\maketitle

\section*{Introduction}

The past decade witnessed an extremely fast development in the studies on non-linear quantum phenomena of light-matter quasiparticles: exciton-polaritons \cite{book1-Kavokin, book2-Sanvitto, book3-Wouters}. Their bosonic nature was widely explored as it enabled for non-equilibrium Bose-Einstein condensation \cite{Kasprzak, Balili_2007, GaN} and the observation of a superfluid behavior \cite{Amo_nature2009, Nardin} with many collective phenomena such as vortices \cite{Lagoudakis_2008}, solitons \cite{Amo_Science2011} and Josephson oscillations \cite{Lagoudakis_PRL2010}. Recent advances in the growth of the high quality semiconductor microcavities with polariton lifetime exceeding their thermalization time leads to the formation of quasi-equilibrium or equilibrium condensates \cite{Caputo_NatMater2018,Sun_PRL2017}.

These effects are very similar to those observed in cold atomic gases. However, one of the main features that distinguishes an exciton-polariton system from a cold atom system is their spin properties. Exciton-polaritons are formed by the strong coupling of excitons in quantum wells with photons in a semiconductor microcavity \cite{Weisbuch}. They inherit their spin degree of freedom from their excitonic component. Excitons have four components with different spin projection on the axis of the quantum well: $J_z = \pm2$, and $J_z = \pm1$. The excitons with spin projection $J_z = \pm2$ remain dark (not active with respect to the coupling to light or photon emission)\cite{Shelykh_2010}. The optically active excitons, with $J_z = \pm1$ spin degeneracy, couple to light forming polaritons with two spin projections. From the point of view of the spin structure, polaritons are therefore bosons with a 1/2 pseudospin.

First reports on the importance to extend the already known physics of degenerate polariton gases to the case with external magnetic field studies come with the theoretical prediction by Y.~G.~Rubo, et al.~\citenum{Rubo_2006}, where a similarity of polariton condensates to superconductors was demonstrated. A full diamagnetic screening of the external magnetic field from the material interior (Meissner effect) and a suppression of superfluidity was predicted. Few experimental works followed this prediction \cite{Kulakovskii, Larionov, Walker, Kochereshko_SciRep2016}, but the polariton Zeeman splitting in typical III-V or II-VI based microcavities is very low, of approx. 100~$\mu$eV at 5~T, being within the range of polariton emission linewidth, which makes the interpretation of the results very difficult.

We have addressed the problem of spinor non-equilibrium polariton condensates in magnetic field with a new technological approach. To observe significant effects of external magnetic field, we enhance the magnetic response via the exchange interaction between the excitonic component of polaritons and magnetic ions placed inside the quantum wells. The interaction of magnetic moments with excitons can lead to a wide range of effects such as giant Zeeman splitting \cite{Komarov_1977} or giant Faraday rotation\cite{Gaj_1978}. In our sample, the cavity structure and Bragg reflectors are free of magnetic ions. Therefore the magnetic interaction occurs only for excitons and the cavity photons are affected neither by Mn ions, nor by the external magnetic field. Such polaritons that mix the properties of cavity polaritons and semimagnetic excitons are called semimagnetic exciton-polaritons.

This work is devoted to the investigation of a non-equilibrium spinor exciton-polariton condensate. We show that starting from a spin multicomponent (i.e. spin up and spin down) condensate, we can tune it smoothly to a single component one with a spin polarization above 95\%, using independently two external parameters:  excitation power and magnetic field. Furthermore, the spin polarization of the condensate is not damped by an increase of the excitation power as observed in the case of nonmagnetic polaritons \cite{Kulakovskii}, but the spin polarization of the semimagnetic polaritons condensate increases with excitation power. In polarization resolved photoluminescence experiments, we study the spin components of polaritons with increasing excitation power under external magnetic field. Below the condensation threshold, the lower polariton (LP) branch is split into opposite circular polarizations due to the giant Zeeman effect. Above the threshold, we observe a single energy condensate which an elliptical polarization is investigated in magnetic field.

\section*{Results}

\begin{figure}[h]
\centering
\includegraphics[width=.666\paperwidth]{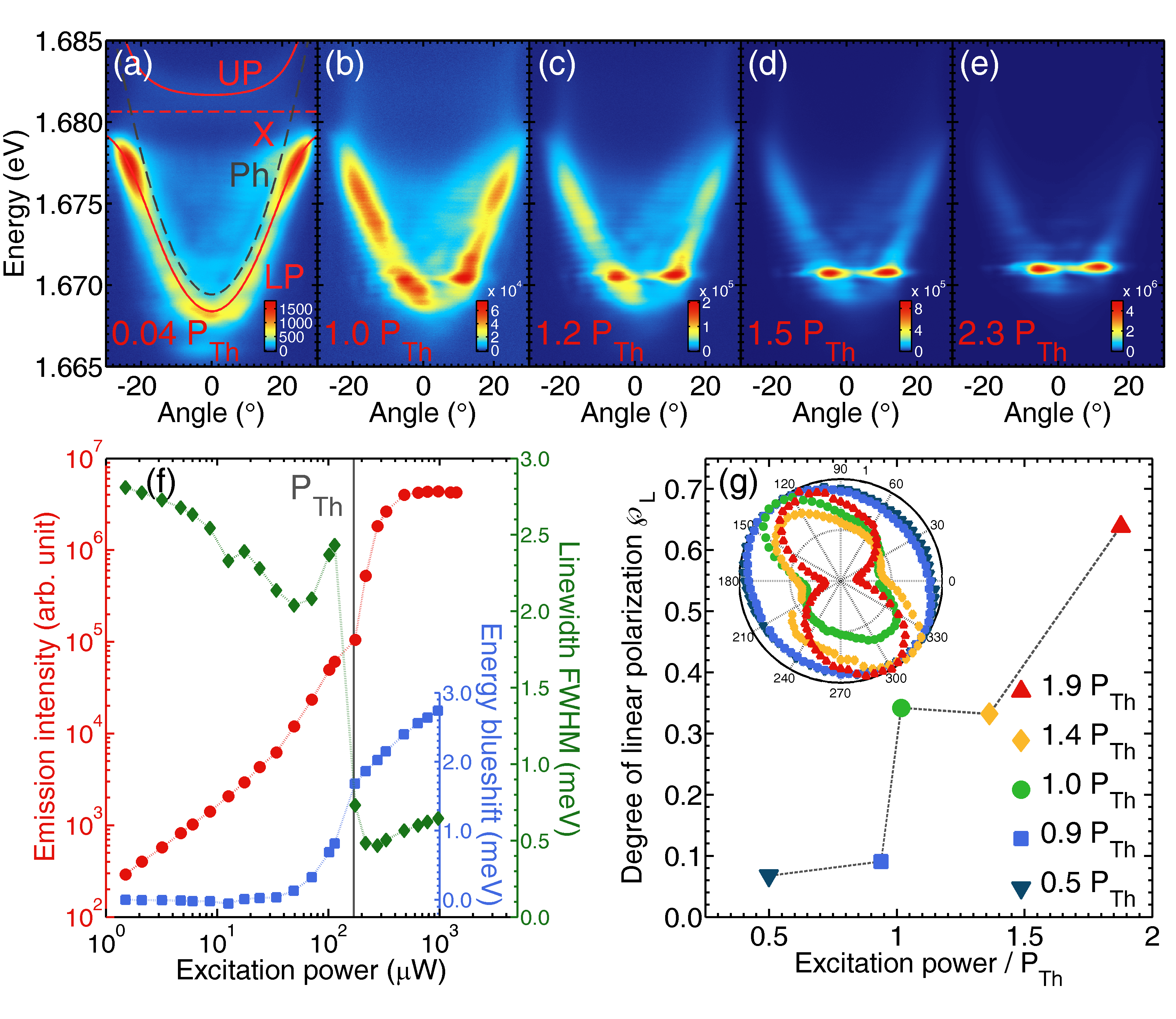}
\caption{\textbf{Exciton-polariton condensation in zero magnetic field.} (a--e) Polariton condensation with increasing non-resonant excitation power presented on angle-resolved photoluminescence maps with corresponding (f) emission intensity, linewidth and energy blueshift 
for the ground state (k=0, zero angle of emission). The condensation threshold power $P_{Th} = 170$~$\mu$W is marked with gray vertical line. (g) Degree of linear polarization of the condensate for different excitation powers. Inset: normalized emission intensity dependence on the angle of linear polarization analyzer. Lines through experimental points are to guide the eye.}
\label{im:Fig1}
\end{figure}

\textbf{Zero field condensation.} 
Non-equilibrium polariton condensation is observed with the increase of the excitation power. Spectra illustrating this effect are presented in Fig.~\ref{im:Fig1}. At low excitation regime, Fig.~\ref{im:Fig1}a, the spectrum is composed of two branches with anti-crossing behavior, typical for exciton-polaritons \cite{Kasprzak, Pietka_PRB2015, Mirek}. The lower (LP) and upper (UP) polariton branches are visible. The two branches are a result of a vacuum field Rabi splitting between the bare exciton, $X$, and cavity photon, $Ph$, resonances. The energy position and dispersion shape of the bare resonances are marked directly in the figure. The Rabi energy at zero magnetic field and close to zero detuning is about 8~meV. 

At low excitation power we observe the accumulation of polaritons at the bottleneck of lower polariton branch. For higher excitation power the population at the bottleneck decreases and the polaritons accumulate at the bottom of lower polariton branch, where the intensity starts to dominate over the intensity at the bottleneck. The non-linear increase in the emission intensity at the bottom of LP is accompanied by an energy blueshift due to polariton--polariton interactions, illustrated in Fig.~\ref{im:Fig1}f, which is a well established signature of polariton condensation \cite{Kasprzak}. Condensation in our structure is observed in localized states, which is a signature of photonic disorder in the CdTe-based samples \cite{Nardin_PRL2009,Pieczarka_PRL2015, Krizhanovskii_PRB2009, Love_PRL2008} and a non-equilibrium character of polariton condensation. The condensate is significantly linearly polarized, as illustrated in Fig.~\ref{im:Fig1}g, similarly to previous reports performed on CdTe structures \cite{Kasprzak,Kasprzak_PRB2007}. The degree of linear polarization increases with the excitation power. The linear polarization of the condensate indicates that in zero magnetic field it has equal densities of both spin components, presumably due to the antiferromagnetic interactions between the spin up and spin down polaritons.

\begin{figure}
\centering
\includegraphics[width=.666\paperwidth]{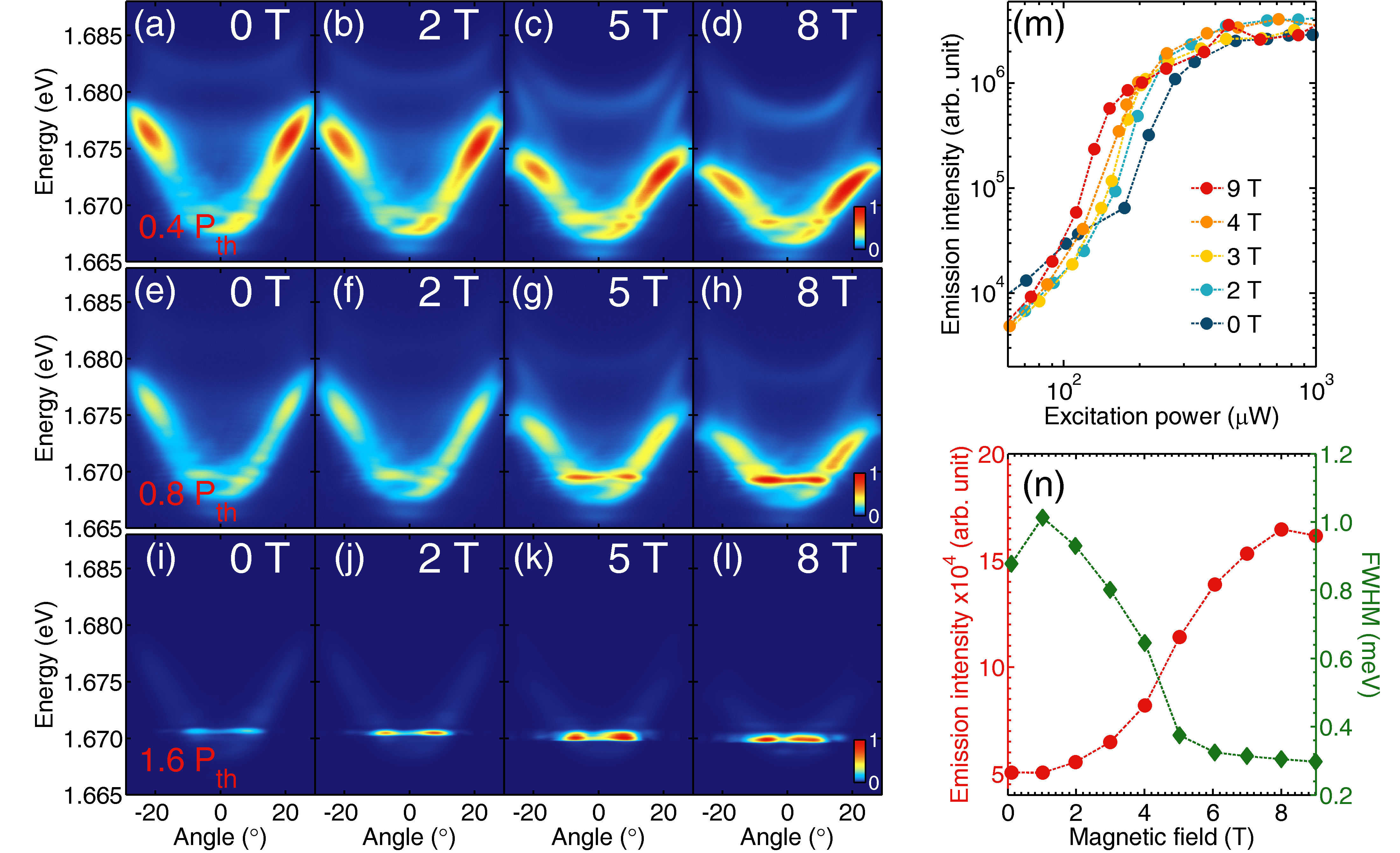}
\caption{\textbf{Exciton-polariton condensation in external magnetic field.} Angle-resolved photoluminescence maps in increasing magnetic field for excitation powers: (a--d) 0.4~$P_{Th}$, (e--h) 0.8~$P_{Th}$ and (i--l) 1.6~$P_{Th}$, where $P_{Th}$ is the condensation threshold power at zero magnetic field, $P_{th}=170$~$\mu$W. Dotted lines marks the position of polariton branches. (m) Dependence of the emission intensity on the excitation power in various magnetic fields. (n) Emission intensity and linewidth as a function of magnetic field for data presented in panels (e--h). Lines through experimental points are a guide to the eye.}
\label{im:Fig2}
\end{figure}

\textbf{Giant Zeeman splitting.}
To split the two spin components of semimagnetic polariton condensate we apply an external magnetic field. The photoluminescence spectra in increasing magnetic field for three different excitation powers are illustrated in Fig.~\ref{im:Fig2}.

The increase of magnetic field (Fig.~\ref{im:Fig2}) results in a splitting of the lower (LP) and upper (UP) polariton branches each into two components, as a direct result of the giant Zeeman splitting of excitons. The magnitude of the observed LP polariton Zeeman splitting is dependent on the exciton--photon detuning, as well as the emission angle, being larger for more exciton-like polaritons (high emission angles) and lower for photon-like polaritons (zero emission angle). This effect was already discussed in Ref.~\citenum{Mirek}, together with negligible diamagnetic shift of semimagnetic excitons. At low excitation power, in Fig.~\ref{im:Fig2}a--d, the Zeeman-splitting is directly visible as a red-shift of LP branch as this component is highly populated. The blue-shifted branch is also visible in Fig.~\ref{im:Fig2}a--d, but its intensity is much weaker. 

With the increase of magnetic field we observe also that the PL intensity from the bottom of the red-shifted LP branch increases. In Fig.~\ref{im:Fig2}e--h, the system in zero magnetic field is slightly below the condensation threshold, but with the increase of magnetic field the ground state starts to macroscopically populate. The condensation is therefore favored in magnetic field, even though the excitation power is kept constant. This process is accompanied by the linewidth narrowing presented in Fig.~\ref{im:Fig2}n. In Fig.~\ref{im:Fig2}m we plot the emission intensity from the bottom of LP branch with increasing excitation power at different magnetic fields. This clearly confirms the decrease of the threshold power in magnetic field discussed in detail in Ref.~\citenum{Rousset_PRB2017} and interpreted as being due to the change of the exciton-polariton detuning and the decrease of the density of states due to the exciton Zeeman splitting. 

In Fig.~\ref{im:Fig2}i--l the excitation power was constant and high enough to create the condensate already at zero magnetic field. With the increase of magnetic field the condensate is more and more populated. Its energy decreases, which is a combined effect of polariton Zeeman splitting (redshift) and the polariton-polariton interactions (blueshift), which is discussed in detail below. At very high magnetic field and at high excitation power (not shown) the condensate emission line broadens, which indicates that the polariton-polariton interactions are strong enough to perturb the condensate and its temporal coherence decreases.

\begin{figure}
\centering
\includegraphics[width=.666\paperwidth]{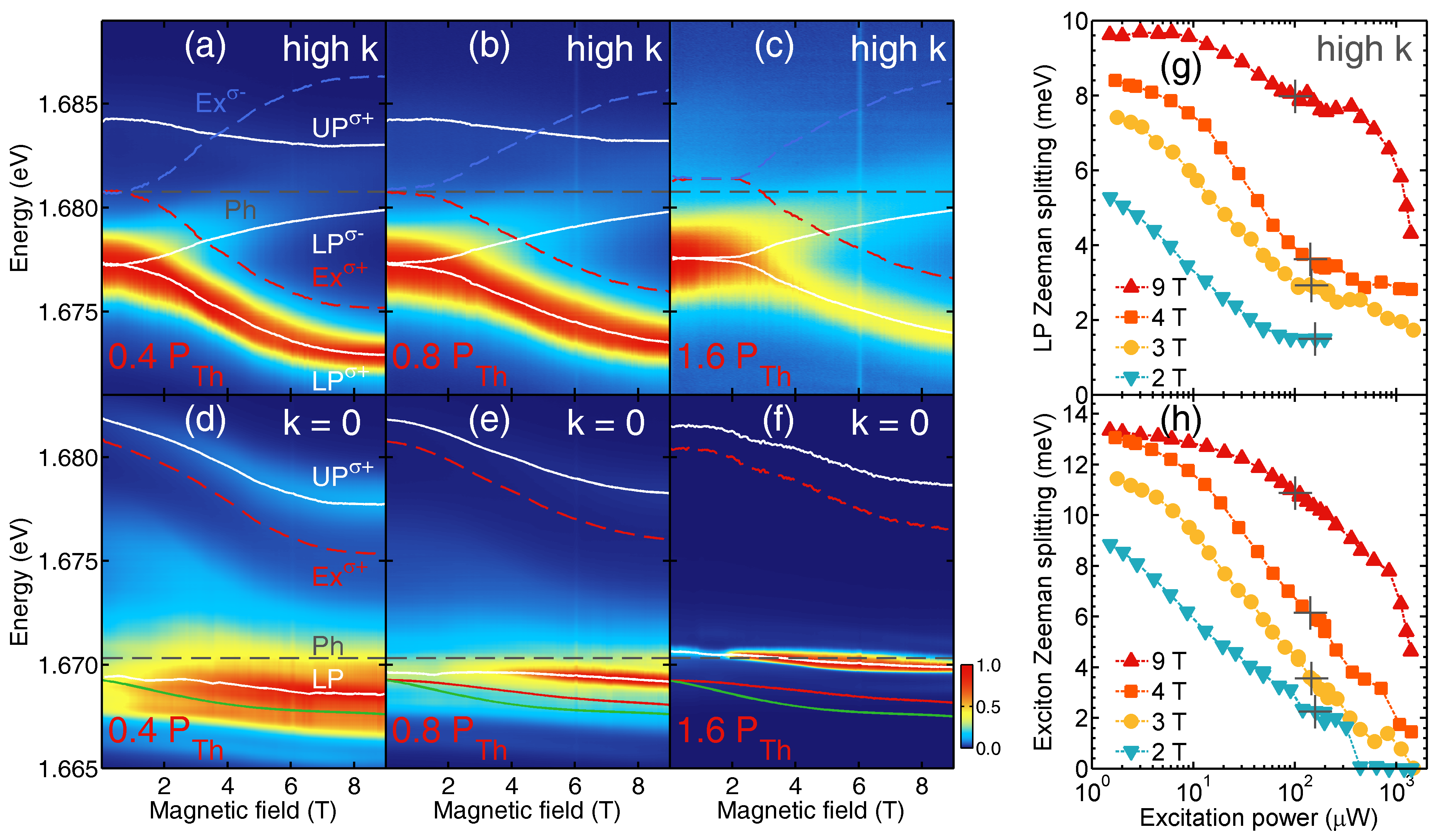}
\caption{\textbf{Reduction of Zeeman splitting in presence of photo-created electron--hole pairs.} Cross-section of the photoluminescence maps from Fig.~\ref{im:Fig2} in magnetic field for three excitation powers: (a,d) 0.4~$P_{Th}$, (b,e) 0.8~$P_{Th}$ and (c,f) 1.6~$P_{Th}$. Panels a--c present the cross-sections at high emission angle, 25$^\circ$, and panels d--f for close to zero emission angle ($\pm10^\circ$). White solid lines depict exciton-polariton emission line energies obtained by fitting the dispersion model to experimental data. Bare states energies are plotted with dashed lines: gray for photon and red/blue for $\sigma^+/\sigma^-$ polarized excitons. Red solid lines for higher excitation powers present LP energy expected from UP emission line energy. $P_{Th}$ is condensation threshold power at zero magnetic field equal to 170~$\mu$W. (g) LP polariton Zeeman splitting at 25$^\circ$ emission angle and (h) bare exciton Zeeman splitting at 2, 3, 4 and 9~T for increasing excitation power. Gray cross marks in (g--h) show condensation threshold power for each magnetic field. Green solid lines give a prediction of LP energy at extremely low excitation power (3~$\mu$W), where the maximum value of the exciton Zeeman splitting is expected. Lines through experimental points are to guide the eye.}
\label{im:Fig3}
\end{figure}

Let us now discuss the effect of polariton-polariton interaction on the energy shift of semimagnetic polariton condensate in magnetic field. This will be done by comparing the LP energy shift in the linear, Fig.~\ref{im:Fig3}d; and the non-linear (Fig.~\ref{im:Fig3}e,f) regimes. The magnitude of the exciton Zeeman splitting can be traced at high emission angles, where the LP branch is strongly exciton-like, as illustrated in Fig.~\ref{im:Fig3}a--c. The energy of the polariton branch is marked with solid white lines, the relevant bare $J_z = +1$/$J_z = -1$ exciton energies are presented with red/blue dashed lines, respectively, and photon energy with grey dashed line. 

We observe in Fig.~\ref{im:Fig3}a--c that at higher excitation power Zeeman splitting of LP polaritons (and consequently the exciton Zeeman splitting) is reduced. To study this behavior in detail we increased the excitation power at various magnetic fields as presented in Fig.~\ref{im:Fig3}g,h. We directly relate the reduction of Zeeman splitting (Fig.~\ref{im:Fig3}g,h) to the depolarization of the magnetic ion system caused by interactions with a large number of electrons and holes created at high excitation power. This effect is widely known for diluted magnetic semiconductors and the dependence on magnetic field of the exciton energy can be described by a Brillouin function with increased effective temperature\cite{Koenig,Golnik_2004}. Nonetheless, the Zeeman splitting is not completely quenched even when the excitation exceeds the condensation threshold power. The threshold power is marked with crosses in Fig.~\ref{im:Fig3}g,h and the polariton Zeeman splitting can be as high as 8~meV at 9~T at the threshold.

The above analysis of the Zeeman splitting and the change of detuning caused by the energy shift of the excitonic component allowed us to plot the expected LP energy at zero emission angle in the magnetic field at which condensation occurs, see Fig.~\ref{im:Fig3}d--f. Green solid lines show the prediction of LP energy at extremely low excitation power, where the maximum value of exciton Zeeman splitting is expected. Red solid lines show the prediction of LP energy in magnetic field including the effect of Mn-ions depolarization. At higher excitation powers the red line illustrates smaller shifts. The white line marks the directly measured energy of the polariton condensate. The energy of the condensate (white line) differs significantly from the prediction (red line) suggesting that nonlinear interactions between polaritons in the condensate significantly shifts its energy in magnetic field, moreover the higher the excitation power, the stronger these effects. The redshift of the condensate in magnetic field is still observed, which proves that the system is still in the strong coupling regime, as only the excitonic component is affected by magnetic field.

\begin{figure}
\centering
\includegraphics[width=.666\paperwidth]{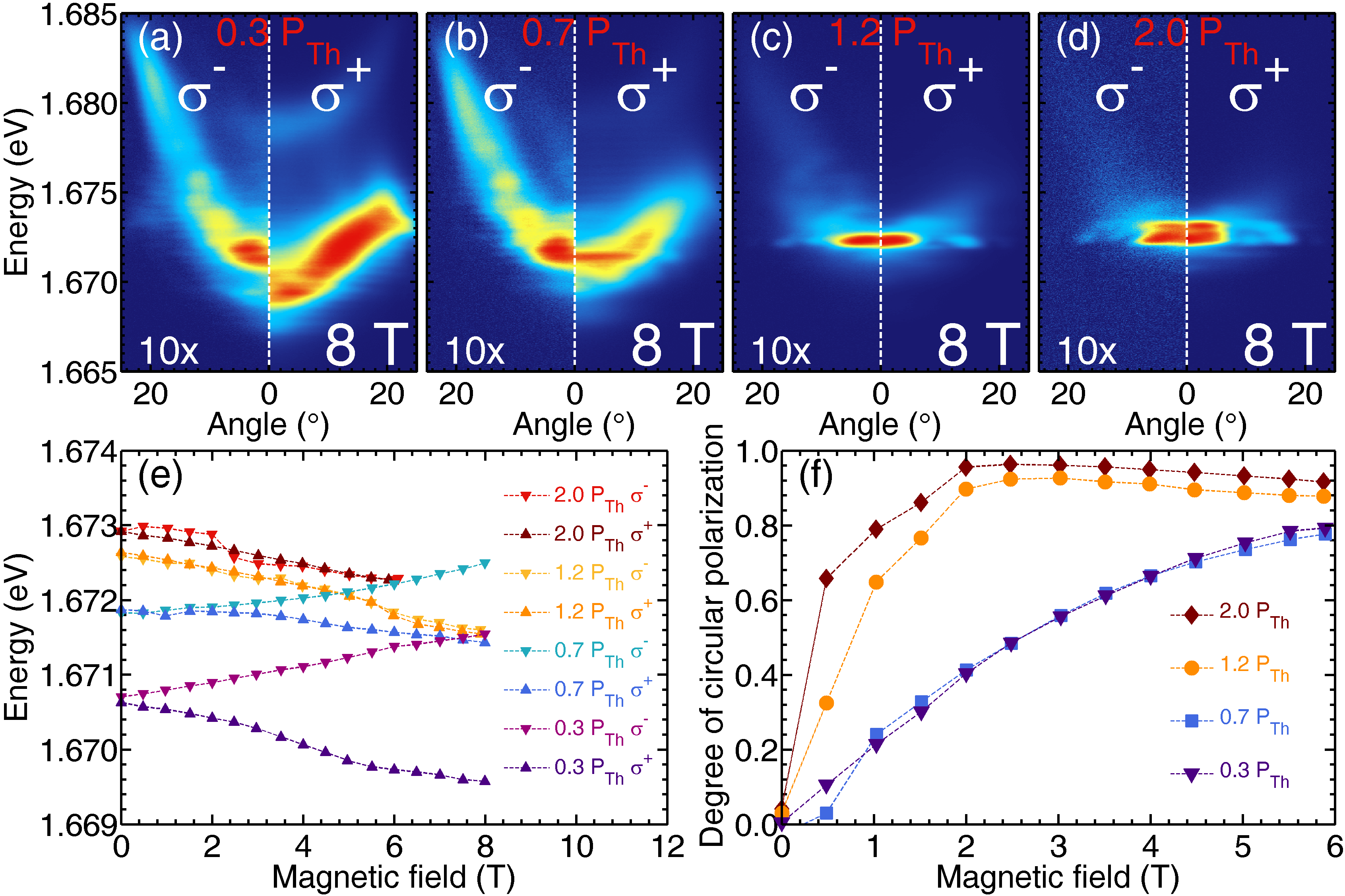}
\caption{\textbf{Spin properties of exciton-polariton condensate.} (a--d) Angle resolved PL maps detected in both circular polarizations at 8~T for excitation powers: (a) 0.3~$P_{Th}$, (b) 0.7~$P_{Th}$, (c) 1.2~$P_{Th}$ and (d) 2.0~$P_{Th}$. Emission intensity in weaker $\sigma^-$ polarization is multiplied 10 times. (e) Energies of $\sigma^+$ and $\sigma^-$ emission lines at the LP ground state at zero emission angle in function of magnetic field. (f) Degree of circular polarization of the LP emission as a function of magnetic field.}
\label{im:Fig4}
\end{figure}

\textbf{Spin polarization of polariton condensate.}
In order to plot the spin polarization of the condensate we performed measurement resolved in circular polarizations. Photoluminescence maps at 8~T for four excitation powers are presented in Fig.~\ref{im:Fig4}a--d.  Below the condensation threshold the LP branch energy is split into two components with opposite circular polarizations, see Fig.~\ref{im:Fig4}a,b. In contrast, above threshold as shown in Fig.~\ref{im:Fig4}c,d the condensate appears as a single energy state in both $\sigma^+$ and $\sigma^-$ polarizations. Only at high condensate population we observe the appearance of the second line in the spectra (Fig. 4d) that we believe is due to the non-equilibrium character of our condensate. The energy dependence of the LP polariton branch at zero emission angle on magnetic field is depicted in Fig.~\ref{im:Fig4}e. At zero magnetic field, the difference in emission energy between results for different excitation powers is due to polariton-reservoir and polariton-polariton interactions.

At the lowest excitation power, the energy of the emission line is dominated by the Zeeman splitting with no diamagnetic shift observable. For stronger excitation but still below the condensation threshold, the behavior is similar, but with lower value of Zeeman splitting. However, above the condensation threshold the situation changes significantly. The condensate appears in both circular polarizations with the same energy, and its energy decreases in magnetic field. Due to the interplay of the Zeeman effect and polariton-polariton interactions, the condensate ground state in magnetic field is elliptically polarized and can be observed in both $\sigma^+$ and $\sigma^-$ polarization, with the resulting degree of circular polarization presented in Fig.~\ref{im:Fig4}f. Almost 90\% of $\sigma^+$ spin polarization is reached at around 3~T. The higher the excitation power, the faster the build-up of spin polarization in magnetic field. This effect demonstrates that stimulated scattering to the condensate is spin dependent and once the imbalance between spin-up and spin-down polariton densities appears, it is further increased. The spin polarization of excitonic reservoir in external magnetic field plays also an important role, as it favors only one spin component.

Condensate polarization properties in magnetic field are also summarized in Supplementary Information~\cite{SI}. A highly linearly polarized condensate at 0~T changes into an elliptically polarized condensate after applying external magnetic field. 

\section*{Discussion}

As compared to the theoretical prediction,\cite{Shelykh_2010} we qualitatively reproduce the full diagram of spin polarization of the condensate as a function of magnetic field and condensate population. The polarization of the condensate evolves from linear to circular as predicted. Concerning the energy components of the condensate, we observe a single energy state with no additional component in the spectra for higher field values \cite{Rubo_2006, Sturm}. We should stress however that our system is out of thermodynamic equilibrium and the observed effect can differ significantly from those predicted for thermal equilibrium.

It is worth to notice that there is a qualitative difference between circular polarization reported here for semimagnetic polaritons and previously reported studies of the magnetic field effect on the polarization of polariton condensates in GaAs based microcavities\cite{Kulakovskii}. In previous reports, even in magnetic field, the degree of circular polarization was decreasing with polariton density, so stimulation to spin polarized state was not effective enough. Here, we show that for semimagnetic polaritons the degree of circular polarization is increasing with magnetic field starting from the lowest magnetic field measured and then it still increases with polariton density (Fig. 4f). This proofs condensation in spin polarized state.

\section*{Methods}
Our sample was grown by molecular beam epitaxy. Microcavity consists of two Bragg mirrors (20 pairs on top and 23 on the bottom) of Cd$_{0.77}$Zn$_{0.13}$Mg$_{0.10}$Te for the high ($n_{high}=2.97$) and Cd$_{0.43}$Zn$_{0.07}$Mg$_{0.50}$Te for the low $n_{low}=2.61$ refractive index. In $3\lambda$-thick cavity from high refractive index material four Cd$_{0.83}$Zn$_{0.16}$Mn$_{0.01}$Te 20~nm wide quantum wells are placed in the anti-nodes of the electric field.  Less than 1$\%$ manganese concentration (typically 0.5$\%$) is present only in the quantum wells\cite{Rousset_JCG2013,Rousset_APL2015}. Due to the low Mn concentration the quantum well is paramagnetic. Strong light-matter coupling in this structure and the giant Zeeman effect of cavity polaritons in magnetic field is described in \citenum{Mirek}.

Sample was excited non-resonantly with a ps laser pulses. The non-resonant excitation was chosen to create an excitonic reservoir composed of excitons with high energy. The time interval between the ps pulses assured reduced heating of the crystal lattice.  The energy of the excitation equal to 1.746~eV corresponds to the first energy minimum observed in the reflectivity spectra of the sample. Angle-resolved photoluminescence spectra was collected from the sample kept at 4.5~K in a superconducting magnet with magnetic field in Faraday configuration up to 9~T. 

\textbf{Data availability.} The datasets generated and analysed during the current study are available from the corresponding author on reasonable request.

\section*{Acknowledgements}

This is a post-peer-review, pre-copyedit version of an article published in \textit{Scientific Reports}. The final authenticated version is available online at: \url{http://dx.doi.org/10.1038/s41598-018-25018-2}.

This work was supported by the National Science Center, Poland under
projects 2014/13/N/ST3/03763, \linebreak 2015/16/T/ST3/00506, 2015/18/E/ST3/00558,
2015/18/E/ST3/00559, and 2015/17/B/ST3/02273. This
study was carried out with the use of CePT, CeZaMat and NLTK
infrastructures financed by the European Union - the European Regional
Development Fund. Scientific work co-financed from the Ministry of Higher
Education budget for education as a research project "Diamentowy Grant":
0010/DIA/2016/45 and 0109/DIA/2015/44. The authors wish to thank Maria Vladimirova for valuable discussions.

\section*{Author contributions statement}
M. K., R. M., K. L., D. S., J. S., B. P. conducted optical experiments, J.-G. R. and W. P. grew and characterized the sample,
M. M. provided theoretical input, M. N. took part in analysis of magneto-optical effects, B. P. prepared final version of the manuscript. All authors discussed the results and reviewed the manuscript.

\end{document}


\thispagestyle{empty}

{\raggedright\sffamily\bfseries\fontsize{20}{25}\selectfont Supplementary Information for Spin polarized semimagnetic exciton-polariton condensate in magnetic field\par}%
\vskip10pt
{\raggedright\sffamily\bfseries\fontsize{12}{12}\selectfont  Mateusz Kr\'ol$^1$, Rafa\l~Mirek$^1$, Katarzyna Lekenta$^1$, Jean-Guy Rousset$^1$, Daniel Stephan$^1$,
Micha\l~Nawrocki$^1$, Micha\l~Matuszewski$^2$, Jacek Szczytko$^1$, Wojciech Pacuski$^1$, and
Barbara Pi\k{e}tka$^{1,*}$\par}
\vskip10pt
{\raggedright\sffamily\fontsize{10}{12}\selectfont  $^1$Institute of Experimental Physics, Faculty of Physics, University of Warsaw, ul.~Pasteura 5, PL-02-093 Warsaw, Poland\\
$^2$Institute of Physics, Polish Academy of Sciences, al.~Lotnik\'{o}w 32/46, PL-02-668 Warsaw, Poland\\
$^*$barbara.pietka@fuw.edu.pl\par}

\begin{figure}[h]
\centering
\includegraphics[width=.666\paperwidth]{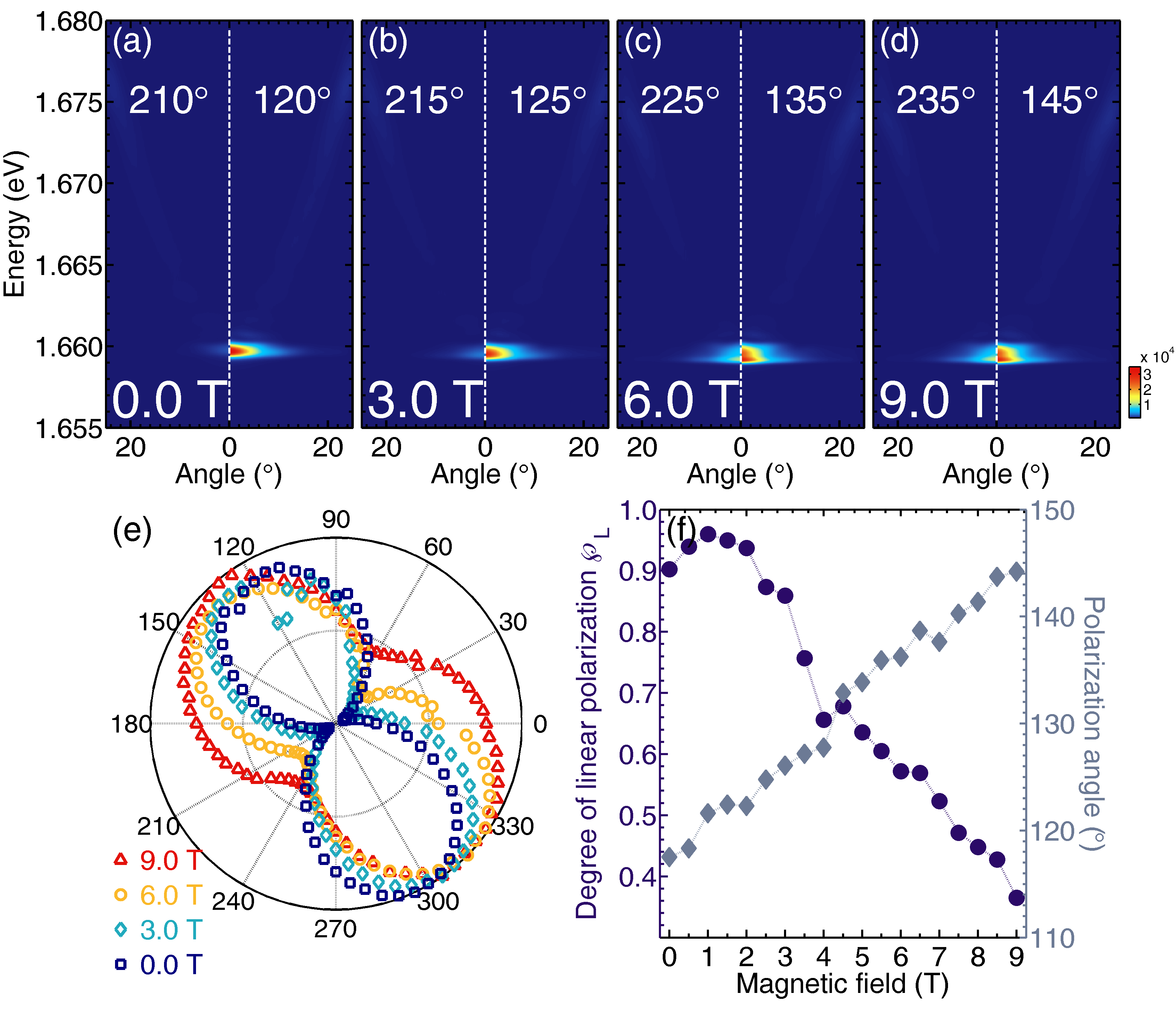}

\caption{\textbf{Polarization properties of the condensate in external
magnetic field.} A highly linearly polarized condensate at 0~T, changes to elliptically-polarized condensate after applying external magnetic field. 
(a--d) Angle resolved PL maps detected in two
perpendicular linear polarizations for increasing magnetic field at
constant excitation power $P_{ex} = 1.5$~$P_{Th}$. (e) Normalized
condensate emission intensity dependence on the angle of the linear
polarization analyzer. (f) Degree of linear polarization and angle of the
linear polarization plane as a function of magnetic field. Lines through
experimental points are to guide the eye. The angle of the linear polarization plane rotates by 27$^\circ$ between 0 and 9~T. This effect we assign to the Faraday rotation in the quantum wells, which is additionally enhanced by the multiple reflections inside the cavity.}
\label{im:Fig5}
\end{figure}